\documentclass[12pt,tightenlines,aps,prc,amsfonts,preprintnumbers,superscriptaddress,showpacs,nofootinbib]{revtex4}

\usepackage{epsfig,epsf,graphics,psfrag}
\usepackage{color}
\usepackage{amsfonts,amssymb,stmaryrd,latexsym,amsmath}
\usepackage[english]{babel}

\usepackage{amsmath}
\usepackage{booktabs}
\usepackage{multirow}
\usepackage{units}
\usepackage{array}
\usepackage{braket}

\newcommand{\beq}{\begin{equation}}
\newcommand{\eeq}{\end{equation}}
\newcommand{\beqa}{\begin{eqnarray}}
\newcommand{\eeqa}{\end{eqnarray}}



\begin{document} 

\title{Scattering cluster wave functions on the lattice  using the adiabatic projection method}

\author{Alexander Rokash}
\email{alexander.rokash@ruhr-uni-bochum.de }
\affiliation{Institut f\"{u}r Theoretische Physik II, Ruhr-Universit\"{a}t Bochum, 44870 Bochum, Germany}
\author{Michelle Pine}
\email{mjmantoo@ncsu.edu}
\affiliation{Department of Physics, North Carolina State University,
        Raleigh, North Carolina 27695, USA}
\author{Serdar Elhatisari}
\email{selhati@ncsu.edu}
\affiliation{Helmholtz-Institut f\"ur Strahlen- und Kernphysik (Theorie) and
        Bethe Center for Theoretical Physics, Universit\"at Bonn, 53115 Bonn, Germany}
\affiliation{Department of Physics, North Carolina State University,
        Raleigh, North Carolina 27695, USA}
\author{Dean Lee}
\email{dean_lee@ncsu.edu}
\affiliation{Department of Physics, North Carolina State University,
        Raleigh, North Carolina 27695, USA}
\author{Evgeny Epelbaum}
\email{evgeny.epelbaum@ruhr-uni-bochum.de}
\affiliation{Institut f\"{u}r Theoretische Physik II, Ruhr-Universit\"{a}t Bochum, 44870 Bochum, Germany}
\author{Hermann Krebs}
\email{hermann.krebs@ruhr-uni-bochum.de} 
\affiliation{Institut f\"{u}r Theoretische Physik II, Ruhr-Universit\"{a}t Bochum, 44870 Bochum, Germany}

\date{\today}

\begin{abstract}
The adiabatic projection method is a general framework for studying scattering
and reactions on the lattice. It provides a low-energy effective
theory for clusters which becomes exact in the limit of large
Euclidean projection time. Previous studies have used the adiabatic
projection method to extract scattering phase shifts from
finite periodic-box energy levels using L{\"u}scher's method.  In this paper
we demonstrate that scattering observables can be computed directly
from asymptotic cluster wave functions. For a variety of examples in
one and three spatial dimensions, 
we extract elastic phase shifts from asymptotic cluster standing waves
corresponding to spherical wall boundary conditions.  We find
that this approach of extracting scattering wave functions from the adiabatic Hamiltonian to be less sensitive to small stochastic and
systematic errors as compared with using  periodic-box energy levels.

\end{abstract}


\maketitle

\section{Introduction}
\emph{Ab initio} description of scattering and reactions involving
nuclei is one of the major challenges in computational nuclear
physics. Recent progress along this line has been achieved using
resonating group methods
\cite{Navratil:2010jn,Navratil:2011zs,Romero-Redondo:2014fya},
fermionic molecular dynamics \cite{Neff:2010nm,Neff:2010uk},  the
coupled-cluster expansion, see \cite{Hagen:2013nca} for a review
article, 
and variational and Green's function Monte
Carlo methods \cite{Nollett:2011qf,Carlson:2014vla}.  For calculations involving lattice methods, there has
been progress in using finite periodic volumes to analyze
coupled-channel scattering
\cite{Liu:2005kr,Lage:2009zv,Briceno:2013lba,Briceno:2012yi,Briceno:2013bda,Briceno:2014oea,Doring:2012eu}
and three-body systems
\cite{Polejaeva:2012ut,Briceno:2012rv,Meissner:2014dea}.  In this
paper we concentrate on the adiabatic projection method for
calculating nuclear reactions from lattice simulations in the
framework of chiral effective field theory (EFT). See
\cite{Epelbaum:2011md,Epelbaum:2012qn,Lahde:2013uqa,Epelbaum:2013paa}
for some recent results using lattice EFT. The general
strategy in the adiabatic projection formalism as formulated in the
pioneering work \cite{Rupak:2013aue,Pine:2013zja,Elhatisari:2014lka} involves two
steps. First, one uses the Euclidean time projection method to
determine an adiabatic Hamiltonian for the participating nuclei
starting from the microscopic Hamiltonian derived in chiral EFT.
In the second step,  one uses a technique such as L\"uscher's finite-volume method \cite{Luscher:1986pf,Luscher:1990ux} to extract the corresponding scattering phase
shifts \cite{Pine:2013zja}
. 

The adiabatic  projection formalism has been successfully
benchmarked against continuum calculations for fermion-dimer scattering.
However for
heavier systems, L\"uscher's finite-volume energy approach for extracting scattering phase shifts is expected to suffer from potentially large errors due to stochastic and systematic uncertainties in the lattice Monte Carlo
energies. In this paper, we explore various techniques to access
scattering on the lattice that do not require a high-accuracy determination of the
energy spectrum. For simple three-body systems in one and three
dimensions, we demonstrate that scattering phase shifts can be
reliably extracted from the asymptotic cluster wave functions within
the adiabatic projection method.   

Our paper is organized as follows. We begin with introducing the
adiabatic projection method in some detail in section \ref{method}. 
Section \ref{sec:luescher} describes L{\"u}scher's finite volume method, while section \ref{sec:WF} describes the extraction of the asymptotic
cluster wave functions on the lattice. The microscopic Hamiltonian
used in our work is specified in section \ref{sec:H}, which also
provides details on the computation of the adiabatic Hamiltonian. 
Various approaches for extracting the two-cluster elastic scattering phase shifts
in one and three spatial dimensions on the lattice are introduced and applied in
sections \ref{sec:Methods} and \ref{sec:Results},
respectively. Finally, the main results of our
study are summarized in section \ref{sec:Summ}.

\section{The adiabatic projection method}
\label{method}

\begin{figure}[tb]
\includegraphics[scale=0.4]{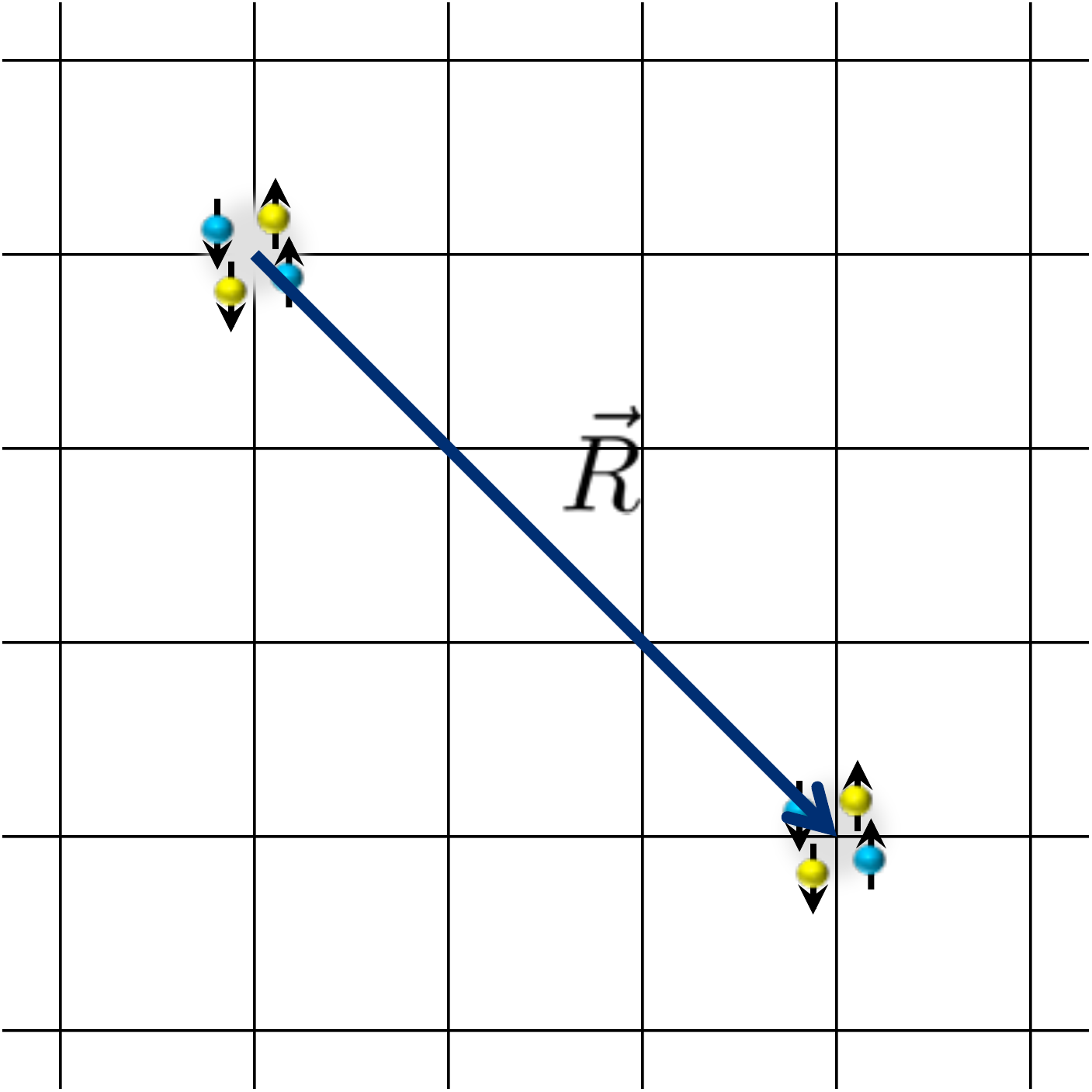}
\caption{(Color online) Two-body cluster initial state $|\vec{R}\rangle$ separated by
  the displacement
vector $\vec{R}$.}
\label{fig:clusters}
\end{figure}

The adiabatic projection method treats the cluster-cluster scattering
problem on the lattice by using Euclidean time projection  to
determine an 
adiabatic Hamiltonian for the participating clusters.  
When the temporal lattice spacing is nonzero, an adiabatic transfer
matrix rather than the Hamiltonian is constructed, but the method is
essentially the same. We start with an $L^3$ periodic 
lattice and set of two-cluster states $|\vec{R}\rangle$ labeled
by their separation vector $\vec{R}$, as illustrated in
Fig.~\ref{fig:clusters}. In general, there are spin and flavor indices
for these states, but we suppress writing the indices for notational
simplicity. The exact form of these two-cluster states is not
important except that they are localized so that for large separations
they factorize as a tensor product of two individual clusters, 
\begin{equation}
|\vec{R}\rangle=\sum_{\vec{r}} |\vec{r}+\vec{R}\rangle_1\otimes|\vec{r}\rangle_2. \label{eqn:single_clusters}\\ 
\end{equation}
   These states are propagated in Euclidean time to form dressed cluster states,
 \begin{align}
\vert \vec{R}\rangle_\tau   =\exp(-H\tau)\vert \vec{R}\rangle .
\end{align}  
By evolving in Euclidean time with the microscopic Hamiltonian, deformations and
polarizations
of the interacting clusters
are incorporated automatically, and we are projecting onto the space of low-energy scattering states in our finite volume. In
the limit of large Euclidean time, these dressed cluster states span the
 low-energy subspace of   two-cluster continuum states.  

We evaluate matrix elements of the microscopic Hamiltonian
with respect to the dressed cluster states,
\begin{align}
\left[H_{\tau}\right]_{\vec{R},\vec{R}'} =\ _{\tau}\langle\vec{R}\vert H \vert\vec{R}'\rangle_{\tau}.
\end{align}
 Since the dressed cluster states  $|\vec{R}\rangle_\tau$
are, in general, not orthogonal, we construct a norm matrix $N_{\tau}$ given by the inner product
\begin{align}
\left[N_{\tau}\right]_{\vec{R},\vec{R}'} =\ _{\tau}\langle\vec{R}\vert\vec{R}'\rangle_{\tau}.
\end{align}
From this a Hermitian adiabatic Hamiltonian matrix can be defined using the inverse square root of the
norm matrix,
\begin{align}
\left[ {H^a_{\tau}} \right]_{\vec{R},\vec{R}'} = \ \sum_{\vec{R}'',\vec{R}'''}
\left[ N_{\tau}^{-1/2} \right]_{\vec{R},\vec{R}''} \left[H_{\tau}\right]_{\vec{R}'',\vec{R}'''}\left[ N_{\tau}^{-1/2} \right]_{\vec{R}''',\vec{R}'}.
\label{eqn:Adiabatic-Hamiltonian}
\end{align}
In the limit of large $\tau$, the spectrum of $H^a_{\tau}$ exactly
reproduces the low-energy finite volume spectrum of the microscropic
Hamiltonian $H$. So for the elastic phase shifts, one can take  the
spectrum of $H^a_{\tau}$ and apply 
the finite-volume scaling analysis developed by L\"uscher \cite{Luscher:1986pf,
Luscher:1990ux}.  

\section{L\"{u}scher's finite-volume method}
\label{sec:luescher}

The L\"uscher method~\cite{Luscher:1986pf,Luscher:1990ux} relates the two-body
scattering states in periodic finite volume to the scattering parameters
in the infinite volume and continuum limits. The main idea behind the derivation
of this method is to use the known asymptotic form of the wave function for
a short range interaction
and the periodicity of the system. The one-dimensional result of this ansatz
is~\cite{Luscher:1986pf}
\beqa
\mathrm{e}^{2i\delta(p)}=\mathrm{e}^{-ipL}\,,
\label{eqn:Luescher-1D}
\eeqa
where $p$ is the relative momenta between two
clusters\footnote{Clusters refer to either a point-like particle or a
  composite particle as a bound state of several particles.},
$\delta(p)$ is the scattering phase shift, and we  assume the total
momentum of the two-cluster system to be zero. In three dimensions, the
situation is more complicated due to breaking of the rotational
invariance by the cubic symmetry of finite periodic box. The
scattering phase shifts are directly related to the momentum via the
formula~\cite{Luscher:1986pf,Luscher:1990ux} 
\begin{equation}
        p \cot\delta_{\ell}(p)
        =
                \frac{1}{\pi\,L}
                S(\eta) \quad \text{for $\ell = 0,1$} \,,
                        \label{eqn:phaseshift-005}
\end{equation}
where $\eta = \left(\frac{L p}{2\pi}\right)^{2}$ and $S(\eta)$ is three-dimensional
zeta function,
\begin{align}
S(\eta)
=
\lim_{\Lambda\to\infty} \left[
\sum_{\vec{n}}^{\Lambda}
\frac
{\theta\left(\Lambda^{2} - \vec{n}^{2}\right)}
{\vec{n}^{2} - \eta}-4\pi\,\Lambda
\right]
\,,
\label{eqn:phaseshift-009}
\end{align}
or, in the exponentially accelerated form~\cite{Luscher:1986pf,Luscher:1990ux,Luu:2011ep},
\beq
S(\eta)
 =
 2\pi^{3/2} e^{\eta}(2\eta -1) 
 +e^{\eta} \, \sum_{\vec{n}}
 \frac{e^{-|\vec{n}|^{2}}}{|\vec{n}|^{2}-\eta}
 -\pi^{3/2} \, \int_{0}^{1} d\lambda\frac{e^{\lambda \eta}}{\lambda^{3/2}}
 \left(4\lambda^{2}\eta^{2} -
 \sum_{\vec{n}}e^{-\pi^{2}|\vec{n}|^{2}/\lambda}\right)
 \,.
 \label{eqn:phaseshift-031}
\eeq

The relation between the relative momentum appearing in
Eq.~(\ref{eqn:phaseshift-005})--(\ref{eqn:phaseshift-031}) and the
finite-volume energies for $\ell= 0$ is given by \cite{Bour:2011ef,Bour:2012hn}
\begin{align}
E(p,L) = \frac{p^{2}}{2 \mu} - B_{1}-B_{2} + \bar{\tau}_{1}(\eta) \Delta E_{1}(L)+
\bar{\tau}_{2}(\eta) \Delta E_{2}(L)\,,
\label{eqn:E&momentum-001}
\end{align}
where $\mu$ is the reduced mass of the system, $B_{i}$ is the binding energy
of the cluster $i=\{1,2\}$ in the infinite volume limit,  $\Delta E_{i}(L)
= E_{i}(L)+B_{i}$ is the finite volume energy shifts of the clusters in the
rest frame, and $\bar{\tau}_{i}(\eta)$ is the topological correction factor to
the energy of the cluster $i$,
\begin{align}
\bar{\tau}(\eta)
= 
\frac{1}{\sum_{\vec{k}}
        \big(\vec{k}^{2} -\eta\big)^{-2}} \sum_{\vec{k}}
\frac{ 
        \sum_{i =1}^{3} \cos\left(2\pi k_{i} \, \alpha \right)}{3\big(\vec{k}^{2}
-\eta\big)^{2}}
\,.
\label{eqn:topologicalphases-029}
\end{align}

In Ref.~\cite{Elhatisari:2014lka}, it was found that for $\ell>0$, the
topological corrections are suppressed by the size of the finite
volume, $\bar{\tau}(\eta) = 1 + \mathcal{O}(1/L)$, so that Eq.~(\ref{eqn:E&momentum-001})
becomes
\begin{align}
E(p,L) = \frac{p^{2}}{2 \mu} + E_{1}(L)+ E_{2}(L)\,.
\label{eqn:E&momentum-002}
\end{align}
To minimize potential errors in the calculations using the L\"uscher's
method, we also take into account the effective mass of the clusters on the
lattice.
 L\"uscher's method is a useful and commonly used tool for calculating
scattering parameters. For examples of recent extensions and
generalizations of L\"uscher's original work see Ref.~\cite{Luu:2011ep,
  Fu:2011xz, Leskovec:2012gb, Briceno:2013lba, Briceno:2012yi,
  Briceno:2013bda, Briceno:2014oea, Doring:2012eu,
  Gockeler:2012yj}. 

The simple elegance of  L\"uscher's method is that
all of the information regarding scattering phase shifts is encoded
into finite-volume energy values.  This simplicity can, however, be a
weakness when applied to scattering processes relevant to
low-energy nuclear physics.  The problem is that the binding energies
of the scattering nuclei can be anywhere from a few MeV to tens or
hundreds of MeV, while the finite-volume scattering energy shifts can be as small
as a few keV.  The problem is even more difficult in lattice QCD
calculations where the rest energy of the nucleons is also part of the
calculations. 

\begin{figure}[tb]
\begin{center}
\includegraphics[scale=1.2]{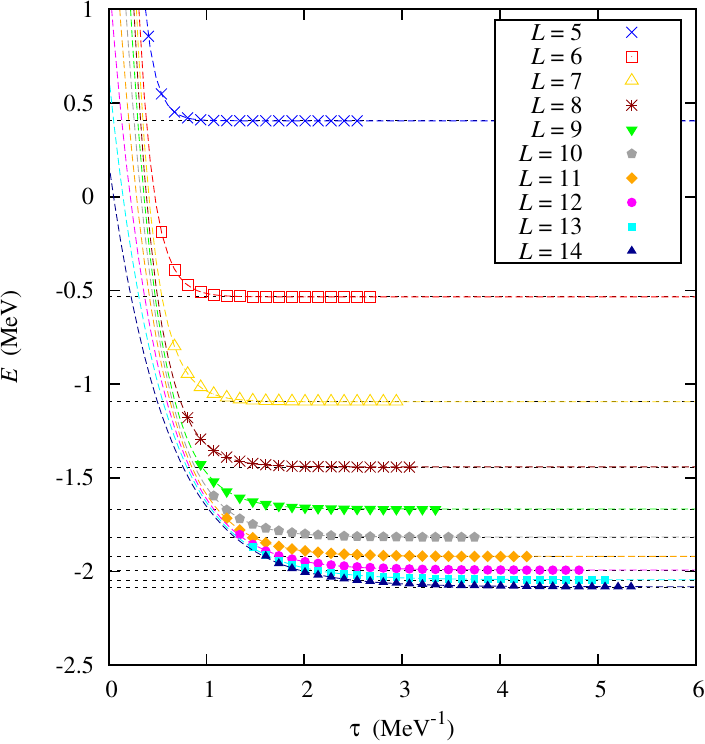}
\caption{(Color online) The colored data points show the lowest energy levels of the
adiabatic Hamiltonian versus projection time $\tau$ for the dimer-fermion system on various $L^3$ periodic lattices. The horizontal black-dashed lines are the lowest energy levels of the microscopic Hamiltonian for the same $L^3$ lattices. The dimer energy is $-2.2246$~MeV and the lattice spacing is 1.97~fm. }
\label{fig:projectionAll_Swave}
\end{center}
\end{figure}

Thus, while the scattering data is encoded in the finite-volume energy
 and waiting to be extracted, the finite-volume energy value is prone
 to several sources of potentially large errors.  In
 Fig.~\ref{fig:projectionAll_Swave}, we show the lowest-lying energy
 state for a dimer-fermion system versus projection time $\tau$ for
 various $L^3$ periodic lattices.  
The dimer energy is set to be 
 $-2.2246~\text{MeV}$ and the lattice spacing is $1.97~\text{fm}$. We
 consider this example in detail later in our discussion. The point we
 emphasize here is that in order to measure the $s$-wave scattering
 phase shift to an error of a few degrees, we need to measure the
 finite-volume energy to an accuracy of about $10~\text{keV}$.  In
 this simple three-particle calculation we are using exact matrix
 methods, and  there are no stochastic errors.  However, in a typical
 large-scale Monte Carlo calculation, there are stochastic errors that
 grow exponentially with projection time $\tau$.  The stochastic errors are
much reduced in a constrained Monte Carlo calculation using, for example,
fixed fermionic nodal constraints.  However here we have a different problem
that the scattering energies may be artificially shifted by the constraints.  In addition to these
 issues, there are also corrections to the binding energies of the
 scattering nuclei due to the finite volume
 \cite{Luscher:1985dn,Konig:2011nz,Konig:2011ti,Bour:2011ef,Bour:2012hn}.      

In view of the problems with finite-volume energy calculations for
low-energy nuclear scattering, we introduce in this paper another
approach for extracting scattering phase shifts which directly
analyzes cluster wave functions generated by the adiabatic
Hamiltonian.  This approach has the advantage of being far less
sensitive to small errors in reproducing the binding energy and
detailed structure of the participating nuclei. 
   
\section{Asymptotic cluster wave functions}
\label{sec:WF}

For each particle of mass $m$, the microscopic Hamiltonian contains a kinetic energy term of the form $-\vec \nabla^2/(2m)$ or, more precisely, the lattice approximation to this operator.   Therefore the Euclidean-time evolution operator $\exp(-H\tau)$ acts as a diffusion operator with diffusion constant inversely proportional to the particle mass $m$.  Since the particles are diffusing in space as a function of Euclidean time, it is necessary to be precise about what we mean by widely-separated clusters at asymptotically large distances.

In order to explain the various time and length scales of our
asymptotic wave function analysis, it is useful to first specify a
relative error tolerance, $\epsilon$, for all steps in our
cluster-cluster scattering calculation.  As we project to large
Euclidean time, any isolated single-cluster initial state will simply relax into the ground state of that cluster system.
We define
$\tau_{\epsilon}$ as the time at which the relative contamination due
to excited cluster states is less than $\epsilon$.

We now consider applying Euclidean time projection for the time duration $\tau_{\epsilon}$ in order to remove excited cluster states. During the time  interval $\tau_{\epsilon}$, each cluster
undergoes spatial diffusion by an average distance proportional to
$\sqrt{\tau_{\epsilon}/M}$, where $M$ is the mass of the cluster. We call this distance the diffusion length. Let
$d_{\epsilon,1}$ be the diffusion length for the first cluster, and
$d_{\epsilon,2}$ be the diffusion length for the second cluster. 
In order to have some widely-separated clusters, we take our periodic box length $L$ to be much larger than
$d_{\epsilon,1}$ and $d_{\epsilon,2}$.  We recall that $\vec{R}$ is the initial separation vector between the two clusters.  When $|\vec{R}| \gg
d_{\epsilon,1},d_{\epsilon,2}$ the dressed cluster state
$|\vec{R}\rangle_{\tau_{\epsilon}}$ consists of non-overlapping
clusters.  We thus can define an asymptotic radius $R_{\epsilon}$ as the
radius such that for $|\vec{R}|>R_{\epsilon}$ the amount of overlap
between the cluster wave packets is less than $\epsilon$. 

In the asymptotic region $|\vec{R}|>R_{\epsilon}$, our dressed
clusters are widely separated and interact only through long range
forces such as the Coulomb interaction.  Therefore,  we can describe
our system in terms of an effective cluster Hamiltonian
$H^{\text{eff}}$ that is simply a free lattice Hamiltonian
for two point particles accompanied by the long-range interactions inherited
from the microscopic Hamiltonian. So in the asymptotic region we
have  
\begin{align}
\left[N_{\tau}\right]_{\vec{R},\vec{R}'} &= c\left[e^{-2H^{\text{eff}}\tau}\right]_{\vec{R},\vec{R}'}, \\
\left[H_{\tau}^ \text{}\right]_{\vec{R},\vec{R}'} 
&= c\left[ e^{-H^{\text{eff}}\tau} H^{\text{eff}} e^{-H^{\text{eff}}\tau}\right]_{\vec{R},\vec{R}'}, 
 \end{align}
where the coefficient $c$ is given by the overlap of the initial
single-cluster states in Eq.~(\ref{eqn:single_clusters}) with the
exact single-cluster energy eigenstates.  Since 
\begin{equation}
\left[N^{-1/2}_{\tau}\right]_{\vec{R},\vec{R}'} = c^{-1/2}\left[e^{H^{\text{eff}}\tau}\right]_{\vec{R},\vec{R}'},
\end{equation}
we conclude that the adiabatic Hamiltonian coincides with the effective cluster Hamiltonian in the asymptotic region,
\begin{equation}
\left[H^a_{\tau}\right]_{\vec{R},\vec{R}'} = \left[H_{}^{\text{eff}}\right]_{\vec{R},\vec{R}'}. \label{adiabatic_equals_eff}
\end{equation}
This last result is quite significant.  We are using Euclidean time
projection to calculate the matrices $N_{\tau}$ and $H_{\tau}$.
This Euclidean time projection entails a considerable amount of
diffusion of the clusters.  However, in the asymptotic region, we are
in essence inverting the diffusion process when computing the
adiabatic Hamiltonian and are left with an effective cluster
Hamiltonian in a position space basis.   

For cases where there are no long range interactions, the scattering
states of the adiabatic Hamiltonian are given by a superposition of Bessel
functions in the asymptotic region.  For the case with Coulomb
interactions, the scattering 
states of the adiabatic Hamiltonian in the asymptotic
region correspond to a superposition of
Coulomb 
wave functions. In the rest of this paper we  extract
cluster-cluster phase shifts for a variety of different examples in
one and three spatial dimensions.  We use spherical hard wall
boundaries placed at some large wall radius $R_{\text{wall}}$ in the
asymptotic region and determine phase shifts by analyzing the
asymptotic scattering wave functions.  As we will show in the following discussion, this method is
robust and accurate even in cases where the corresponding finite-volume
energies produce large errors using L\"uscher's method.    

\section{Lattice Formalism}
\label{sec:H}

\subsection{Microscopic Hamiltonian}

Throughout this work, we consider several systems, all of which are comprised of three point particles. For some cases the particles are distinguishable and in some cases not.  We  specify the quantum statistics explicitly at a later stage. We consider the  limit of where the  range of the interaction is negligible
compared to the scattering length. Then, in the low-energy limit, the interaction
between particles
can be represented by a delta-function.  

Let $b_{s}$ and $b^\dagger_{s}$ be the annihilation and creation operators for each particle species $s$, and let $\rho_{s}(\vec{r})$ be corresponding density operator,
\begin{align}
        \rho_{s}(\vec{r}) =  b_{s}^{\dagger}(\vec{r}) b_{s}(\vec{r})
\,.
        \label{eqn:Latt-densty-op-001}
\end{align}The Hamiltonian in $D$ spatial dimensions has the form $H = H_0 + V$ where
\begin{align}
        H_{0} = - \sum_{s}
        \int d^{D}r \, b_{s}^{\dagger}(\vec{r}) \frac{ \vec{\nabla}^2}{2m_s}
b_{s}(\vec{r}) \,,
        \label{eqn:free-Hamiltonian-001} \\
         V = \sum_{s<s'} C_{s,s'} \int d^{D}r \,
        :\rho_{s}(\vec{r}) \,  \rho_{s'}(\vec{r}): \,,
        \label{eqn:Hamiltonian-001}
\end{align}
and the $::$ symbols indicate normal ordering. On the lattice, we can write the Hamiltonian as $H = H_0 + V$ where
\begin{align}
 {H}_{0} & =\sum_s \sum_{\hat{l}}
        \sum_{\vec{n}}
        \frac{1}{2m_s}b_{s}^{\dagger}(\vec{n})\left[
        2b_{s}(\vec{n})
        -b_{s}(\vec{n}+\hat{l})
        -b_{s}(\vec{n}-\hat{l})
        \right] \,,
        \label{eqn:lattice-Hamiltonian-001} \\
        V &= \sum_{s<s'}\sum_{\vec{n}}C_{s,s'}        
        {:\rho}_s(\vec{n}) \,
        {\rho}_{s'}(\vec{n}):\, .
\end{align}
Here $\hat{l}$ denotes lattice unit vectors along all possible spatial axes.

For $D>1$, a regularization of ultraviolet divergences due to the
zero-range interaction is needed and  provided by the nonzero lattice spacing. We
denote the spatial lattice spacing as $a$. Throughout this paper, we will write all
quantities in lattice units, which are physical units multiplied by
the corresponding power of $a$ in order to render the combination
dimensionless. 

In this paper we consider two examples, the first in one dimension and the second in three dimensions. We have organized our discussion to discuss both examples together, illustrating the methods and results for both examples, one after the other. For both cases the
lattice spacing is taken to be $1.97~\text{fm}$ or $0.0100$ MeV$^{-1}$. For the one-dimensional example, we consider three distinguishable particles with equal masses and calculate the scattering between a particle of species 3 with a dimer composed of particle types 1 and 2. For this case we tune the coupling $C_{1,2}$ to produce a bound state with energy  $-2.0000~\rm{MeV}$.  The couplings $C_{1,3}$ and $C_{2,3}$ are both set equal to exactly one-tenth of the coupling $C_{1,2}$.

For the three-dimensional example, we consider two species of fermions which we label with spins, $\uparrow$ and $\downarrow$.  In this case we calculate scattering between a $\uparrow$ particle and dimer composed of particles $\uparrow$ and $\downarrow$.  The fermion-dimer system corresponds to the
neutron-deuteron scattering in the spin-quartet channel at leading
order of pionless effective field theory. Therefore, we set the
coupling constant $C_{\uparrow,\downarrow}$ to a value for which the dimer
energy has the value of the physical deuteron energy of $-2.2246$ MeV.  

\subsection{Adiabatic Hamiltonian}
We now discuss the implementation of the adiabatic
projection method using a set of initial cluster states as introduced in Eq.~(\ref{eqn:single_clusters}).  The two examples
we consider are a particle-dimer system in one dimension and a fermion-dimer system in three dimensions. For the one-dimensional particle-dimer system, the particle-dimer cluster states are defined as 
\begin{align}
        \ket{\phi_{\vec{R}}} = \sum_{\substack{\vec{n}}}
        b_1^{\dagger}(\vec{n}) b_2^{\dagger}(\vec{n})   b_3^{\dagger}(\vec{n}+\vec{R}) \ket{0} \,.
\end{align}We use vector notation for notational consistency even though there is only one spatial direction. For the three-dimensional fermion-dimer system, we write the fermion-dimer initial cluster states as
\begin{align}
        \ket{\phi_{\vec{R}}} = \sum_{\substack{\vec{n}}}
        b_{\uparrow}^{\dagger}(\vec{n}) b_{\downarrow}^{\dagger}(\vec{n})  b_{\uparrow}^{\dagger}(\vec{n}+\vec{R}) \ket{0},
\end{align}
where $\vec{R}\ne 0$ because of Fermi statistics.

We now project the initial cluster states $\ket{\phi_{\vec{R}}}$ in Euclidean
time $\tau$ using the microscopic Hamiltonian, $H$, for the respective systems.  We use the Trotter approximation,
\begin{align}
        \exp(-{H}\tau) \approx \left(1-a_{t}H \right)^{L_{t}},
        \label{eqn:TrotterApprx}
\end{align}where $a_{t}$ is a time step parameter and $\tau = a_tL_t$. For both the one-dimensional particle-dimer and three-dimensional fermion-dimer systems we have
\begin{align}
        \ket{\phi_{\vec{R}}}_{\tau} = (1-a_{t}H)^{L_{t}} \, \ket{\phi_{\vec{R}}}.
        \label{eqn:Dressed-R}
\end{align}

For large $\tau$ we obtain an accurate representation of the
low-energy spectrum of ${H}$ using the adiabatic Hamiltonian $\left[ {H^a_{\tau}} \right]_{\vec{R},\vec{R}'}$ defined in
Eq.~(\ref{eqn:Adiabatic-Hamiltonian}).

\subsection{Scattering phase shifts from periodic-volume energy levels}

\begin{figure}[tb]
\includegraphics[width=0.47\textwidth]{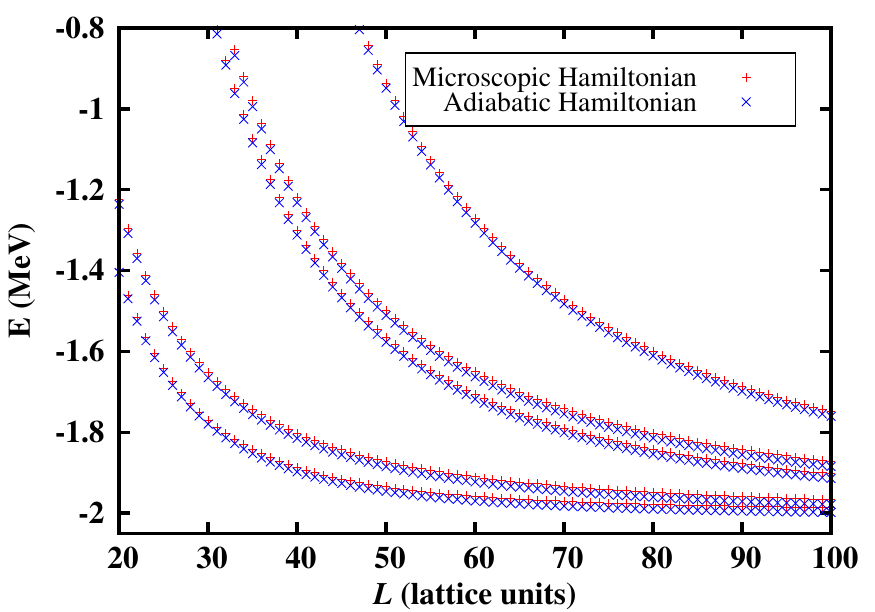}
\hfill
\includegraphics[width=0.47\textwidth]{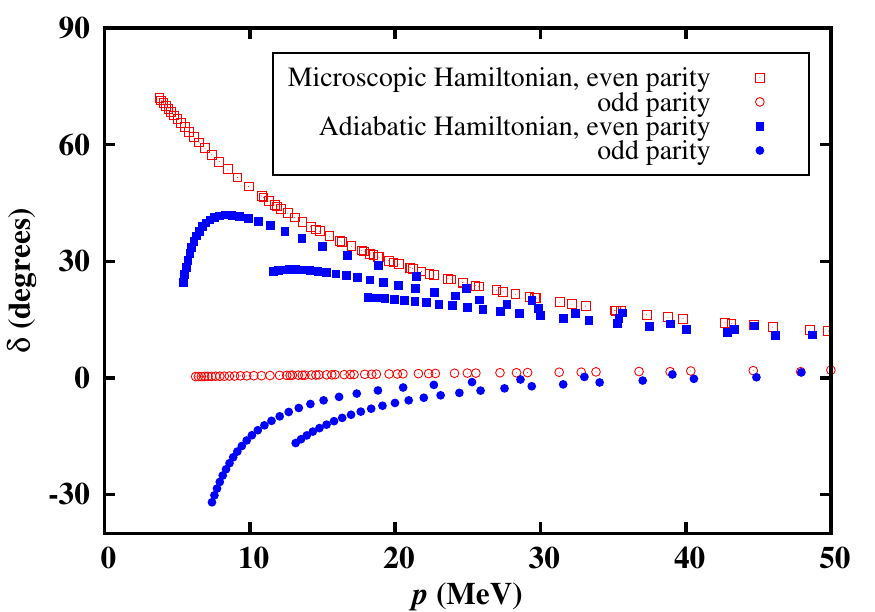}
        \caption{(Color online) Left panel: Finite-volume energies extracted from the
          microscopic Hamiltonian and the two-cluster
          adiabatic Hamiltonian for the particle-dimer system in one
          dimension. Right panel: The particle-dimer scattering phase
          shifts calculated from the data in the left panel using L\"uscher's method. Color online: Red (blue) symbols show the
          results corresponding to the original (adiabatic) Hamiltonian.} 
        \label{pic0}  
\end{figure} 

In Fig.~\ref{pic0}, we compare the energy spectrum of the microscopic Hamiltonian and the two-cluster adiabatic Hamiltonian
for the particle-dimer system in one dimension for $\tau = 0.30~{\rm MeV}^{-1}$. We use the finite-volume energies to calculate the
particle-dimer scattering phase shifts employing L\"uscher's
method. Comparative results for the scattering phase shifts are shown
in the right panel of Fig.~\ref{pic0}.  Although the energy spectra of the adiabatic
Hamiltonian and microscopic Hamiltonian are very similar, the resulting phase shifts have large differences at low
energies. We also see a
disagreement at low energies among phase shifts determined using different adiabatic Hamiltonian energy levels.
This is because low-energy phase shifts are computed using very large
box sizes $L$ where the level spacing is small, and this magnifies any small
discrepancies in the energy values.

We have performed a similar analysis for the fermion-dimer system in three
dimensions. In Table~\ref{table:spectrum_H_Ha}, we compare the
lowest-lying energy states and resulting phase shifts computed using the
microscopic Hamiltonian and adiabatic Hamiltonian for the
dimer-fermion system. The energies $E$ correspond to the microscopic Hamiltonian, while the energies $E(\tau)$ correspond to the adiabatic Hamiltonian. The scattering phase shifts are
calculated using L\"uscher's method. As can be seen from
Table~\ref{table:spectrum_H_Ha}, the accuracy of the finite-volume energies is the range of 40--130 keV.  However the relative error in
the resulting phase shifts becomes as large as 50\%. 
\begin{table}[t]
        \caption{The lowest-lying energy and the corresponding
          scattering phase shifts for the dimer-fermion system computed
          using the L\"uscher's method for various periodic
          lattices. The energies $E$  correspond to the microscopic Hamiltonian while the energies $E(\tau)$ correspond to the  adiabatic Hamiltonian with projection time $\tau$.          } 
        \label{table:spectrum_H_Ha}
\medskip
        \begin{tabular*}{\textwidth}{@{\extracolsep{\fill}}lcclcc}
                \hline\hline
                \multirow{2}{*}{$L$} 
                & \multicolumn{2}{c}{$H$\quad}
                & \multicolumn{3}{c}{$\left[H^a_{\tau}\right]_{\vec{R},\vec{R}'} $}
                \\
                \cmidrule[0.2pt](lr{0.5em}){2-3}
                \cmidrule[0.2pt](lr{0.4em}){4-6}
                &\multicolumn{1}{c}{$E$ {\footnotesize{(MeV)}}}
                &\multicolumn{1}{c}{\quad$\delta$ {\footnotesize{(degrees)}}}
                & \multicolumn{1}{c}{~$\tau$ {\footnotesize{(MeV$^{-1}$)}}}  \quad
                &\multicolumn{1}{r}{$E(\tau)$ {\footnotesize{(MeV)}}}
                &\multicolumn{1}{c}{\quad$\delta$ {\footnotesize{(degrees)}}}
                \\\hline
                ~8~
                & -1.4423319
                & -42.6
                & \quad 0.37
                & -1.4060289 
                & -44.1
                \\
                ~9~
                & -1.6670941
                & -37.9
                &  \quad 0.37
                & -1.6121233
                & -40.9
                \\
                ~10~
                & -1.8171997
                & -33.6
                & \quad 0.34
                & -1.7214154
                & -40.8
                \\
                ~11~
                & -1.9203247
                & -29.8
                &  \quad 0.34
                & -1.8054714
                & -39.6
                \\
                ~12~
                & -1.9929256
                & -26.4
                & \quad 0.34
                & -1.8617182
                & -40.0
                \\\hline\hline   
        \end{tabular*}
\end{table}

Fortunately the adiabatic Hamiltonian contains more usable information than just periodic-lattice energy levels.  As we show in Section~\ref{sec:Methods}, the scattering phase shifts can be determined with far better
accuracy using the properties of the asymptotic scattering wave function.

\section{scattering cluster wave function: methods}
\label{sec:Methods}

Borasoy {\it et al.} introduced a
method to compute phase shifts for point-like
particles on a lattice using a spherical wall boundary
 \cite{Borasoy:2007vy}. As done in the continuum \cite{Carlson:1984zz}, a spherical hard wall of radius $R_{\rm{wall}}$ is imposed on the relative separation of the two
particles. In this study, we consider two-cluster systems, and the spherical
hard wall boundary is imposed on the relative separation of the two
clusters. For two clusters interacting 
via a potential of a finite-range $R$, the wave function at distances
$r>R$  is given by 
\begin{align}
 \Psi_{\ell}^{(p)}(r) = A_{\ell}\cos(pr+\delta_{\ell}-\ell\pi/2) \; & \text{for one dimension},\label{eqn:1D} \\
   \Psi_{\ell,m_{\ell}}^{(p)}(\vec{r}) =
    R_{\ell}^{(p)}(r) \, Y_{\ell,m_{\ell}}(\theta,\phi)   \;    & \text{for three dimensions},
  \label{eqn:Rwall-001}
\end{align}
where $p$ is the relative
momentum of the clusters. For the one-dimensional case, there is no angular momentum, but we nevertheless use the notation $\ell=0$ for even parity and $\ell=1$ for odd parity. In three dimensions, the total wave function
is decomposed into the radial part $R_{\ell}^{(p)}(r)$ and
spherical harmonics $Y_{\ell,m_{\ell}}(\theta,\phi)$. The radial wave function
$R_{\ell}^{(p)}(r)$ has the asymptotic form
\begin{align}
 R_{\ell}^{(p)}(r) = A_{\ell} \, \left[
 \cos\delta_{\ell}(p) 
  \,  j_{\ell}(pr) - \sin\delta_{\ell}(p) \, n_{\ell}(pr) \right] \,,
  \label{eqn:Rwall-005}
\end{align}
where $A_{\ell}$ is a normalization coefficient, and $j_{\ell}$ and
$n_{\ell}$ denote spherical Bessel functions of the first and second
kinds. Therefore, the three dimensional wave function in
Eq.~(\ref{eqn:Rwall-001}) can be rewritten as  
\begin{equation}
 \Psi_{\ell,m_{\ell}}^{(p)}(\vec{r}) = A_{\ell} Y_{\ell,m_{\ell}}(\theta,\phi)
  \left[
 \cos\delta_{\ell}(p) 
   \,  j_{\ell}(pr) - \sin\delta_{\ell}(p) \, n_{\ell}(pr) \right]
   \,.
  \label{eqn:Rwall-006}
\end{equation}

In the asymptotic region, we fit Eq.~(\ref{eqn:Rwall-006}) to the
lattice wave functions emerging from imposing the spherical hard wall at radius
$R_{\rm{wall}}$. We evaluate the spherical harmonics $Y_{\ell,m_{\ell}}(\theta,\phi)$ on the lattice points, noting that there is no exact separation of radial and angular variables on the lattice.  One must take into account the break up of the $2\ell + 1$ spin multiplets according to irreducible representations of the cubic rotation group \cite{Lu:2015gfa,Lu:2014xfa}.   
\begin{table}[tb]
\caption{Irreducible
  representations of the cubic rotation group $\mathrm{SO}(3,Z)$ and relation to spherical
  harmonics for  $\ell \leq 2$.} 
\label{tab:Aofthera&phi}
\medskip
\begin{tabular*}{\textwidth}{@{\extracolsep{\fill}}ccc}
\hline\hline
$\mathrm{SO}(3,Z)$ &  $\, \mathrm{SO}(3) \, $  & $Y^{}_{\ell,m_{\ell}}$ \\[2pt]
\hline
$A_{1}$   &   $\ell=0$ & $\left\{Y_{0,0}\right\}$
\\[2pt]
$T_{1}$   &   $\ell=1$ & $\left\{Y_{1,-1}, \; Y_{1,0}, \; Y_{1,1}\right\}$ 
\\[2pt]
$E_{1}$   &   $\ell=2$ & $\left\{Y_{2,0}, \;  (Y_{2,-2}+Y_{2,2}
                         )/\sqrt{2} \right\}$ 
\\[2pt]
$T_{2}$   &   $\ell=2$ & $\left\{Y_{2,1}, \; (Y_{2,-2}-Y_{2,2})/
                         \sqrt{2}, \; Y_{2,-1}\right\}$
\\[2pt]
\hline\hline
\end{tabular*}
\end{table}
Irreducible representations of the
$\mathrm{SO}(3,Z)$ cubic rotation group are given in
Table~\ref{tab:Aofthera&phi}~for $\ell\le2$ \cite{Johnson:1982yq}.

\begin{figure}[tb]
 \begin{center}
{\includegraphics[width=0.47\textwidth]{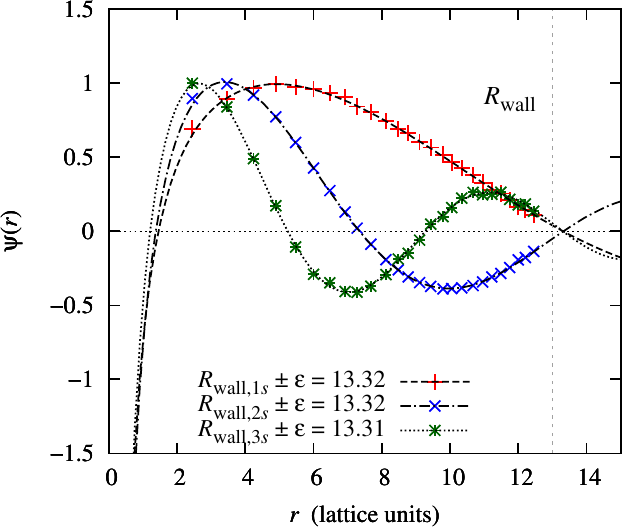}}
\hfill
{\includegraphics[width=0.47\textwidth]{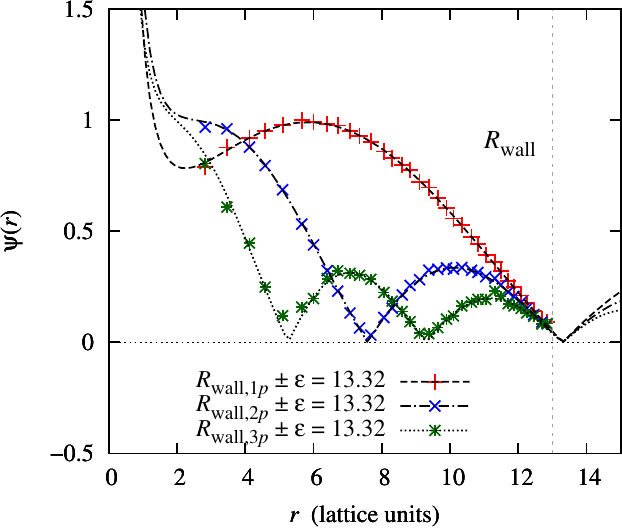}}
 \end{center}
 \caption{(Color online) First three lowest-lying $s$-state wave functions (left panel) and
$p$-wave
    probability distributions (right panel) for the fermion-dimer system in three
dimensions. A spherical hard wall is imposed at $R_{{\rm{wall}}} = 13.0$ lattice units.}%
 \label{fig:spw_func}
\end{figure}

For $\ell=0$, the spherical harmonic $Y_{0,0_{}}(\theta,\phi)$ is
angle-independent, and we can directly match Eq.~(\ref{eqn:Rwall-006})
to the lattice wave functions. However, for $\ell>0$, the angular dependence makes the fitting more difficult. To
resolve this issue, we use the identity 
\begin{equation}
\sum_{m_{\ell}=-\ell}^{\ell} \, \left|Y_{\ell,m_{\ell}}(\theta,\phi)\right|^{2}= \frac{2\ell+1}{4\pi}.
\end{equation}
Since the angular dependence drops out of this expression, it is convenient to work with the wave function probability distribution summed over $m_{\ell}$,
\begin{equation}
\sum_{m_{\ell}=-\ell}^{\ell}\left|\Psi_{\ell,m_{\ell}}^{(p)}(\vec{r})\right|^{2}=\frac{2\ell+1}{4\pi}|A_{\ell}|^2 \left[
 \cos\delta_{\ell}(p) 
   \,  j_{\ell}(pr) - \sin\delta_{\ell}(p) \, n_{\ell}(pr) \right]^2
   \,.
  \label{eqn:Rwall-003}
\end{equation}
To obtain accurate results for the phase shifts within this
approach, it is helpful to address some uncertainties in the
precise location of the wall radius $R_{\rm{wall}}$. While we impose a very large but finite repulsive potential at distances $r \ge R_{\rm{wall}}$, a close examination shows that wavefunction vanishes at some slightly larger radius $R'_{\rm{wall}}=R_{\rm{wall}}+\epsilon$, where $\epsilon$ is some fraction of a lattice spacing.  This is illustrated in
Fig.~\ref{fig:spw_func} for the lowest
three $s$- and $p$-states of the fermion-dimer system in
three dimensions.   
The hard wall potential is imposed for $r \ge R_{{\rm{wall}}}=13.0$ lattice
units.  We find that the first zero of the
1$s$ wave function is at 13.32, the 
second zero of the 2$s$ wave function is at 13.32, and the third zero of the
3$s$ wave function is at 13.31 lattice units. For the first three
lowest $p$-states, we find that corresponding zeros are all at 13.32 lattice units. 

One can extract the scattering phase shifts by doing a three parameter fit of the overall normalization,
momentum and phase shift to the interacting wave functions. We present the results of this fitting procedure later in our discussion. However, we have found more accurate results by making use
of the empirical observation that $R'_{\rm{wall}}$ changes very little when going from the non-interacting
system to the interacting system at approximately the same scattering energy.
We first determine $R'_{\rm{wall}}$ from the lattice wave functions of the non-interacting cluster-cluster
system.
Then, using the same value
of $R'_{{\rm{wall}}}$, we fit the interacting wave functions using a two parameter fit to determine the phase shift of the interacting system using the relations
\begin{align} 
\delta_{\ell}(p) =
  \begin{cases}
   -pR'_{{\rm{wall}}}+\frac{\pi(\ell+1)}{2} \mod \pi &  \quad \text{for one dimension} \\
    \tan^{-1} \left[ \frac{j_{\ell}(pR'_{{\rm{wall}}}/a)}
                          {n_{\ell}(pR'_{{\rm{wall}}}/a)} \right]  & \quad  \text{for three dimensions}\,.
  \end{cases}
\label{eqn:Rwall-009}
\end{align}

\begin{figure}[tb]
 \begin{center}
{\includegraphics[width=0.47\textwidth]{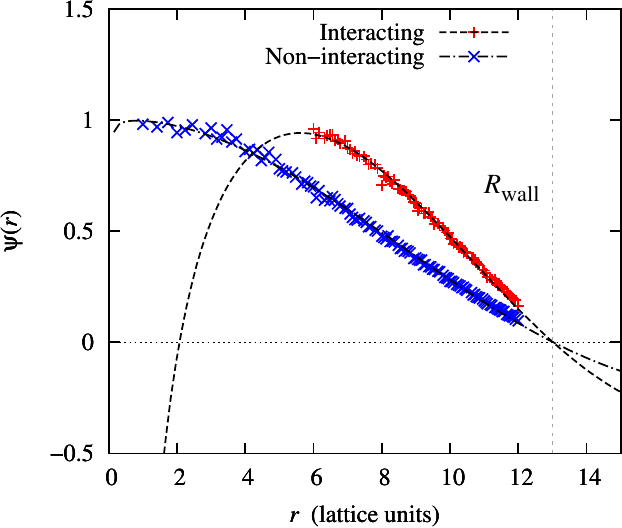}}
\hfill
{\includegraphics[width=0.47\textwidth]{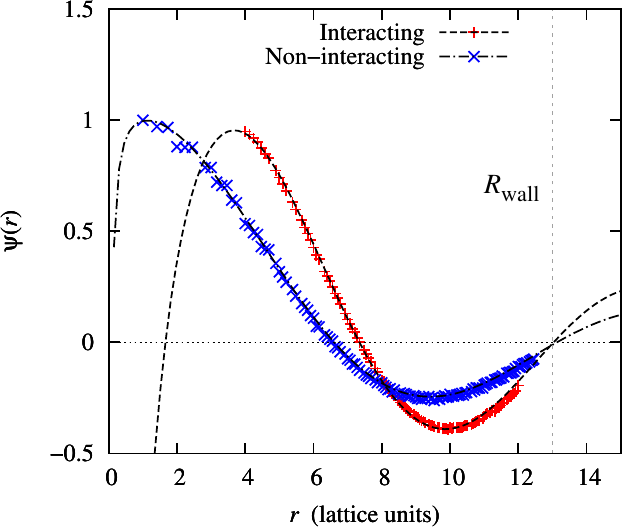}}
 \end{center}
 \caption{(Color online) Matching of the wave functions
   of the 1$s$- (left panel) and
   2$s$-state (right panel). Free and interacting wave
   functions are denoted by (color online: blue) saltires and (color online: red) crosses, respectively.}%
 \label{fig:wf_12swave}
\end{figure}

In Fig.~\ref{fig:wf_12swave}, we show the 1$s$ and 2$s$  non-interacting particle-dimer wave functions used to
calculate  
$R'_{{\rm{wall}}}$ and the corresponding interacting fermion-dimer
wave functions used to determine the $s$-wave scattering phase shift
$\delta_{0}(p)$. 

\section{scattering cluster wave function: Phase shift results}
\label{sec:Results}

We now compute the scattering phase shifts using the adiabatic projection method and scattering cluster wave functions in our one-dimensional particle-dimer system and three-dimensional fermion-dimer system. The results are benchmarked against phase shifts extracted from the exact
three-body energy spectrum obtained using L\"{u}scher's method. For the three-dimensional fermion-dimer system, the
three-body energies are computed using the Lanczos iterative
eigenvector method with a space of $L^{6}$ basis states.  These can be viewed as exact lattice phase shifts.  We note that while the adiabatic projection method calculations can be applied to much larger systems using lattice Monte Carlo, these exact Lanczos calculations are limited to small systems. 

\subsection{Particle-dimer scattering in one dimension}


{\bf }

\begin{figure}[tb]
\includegraphics[angle=0,width=0.38\textwidth]{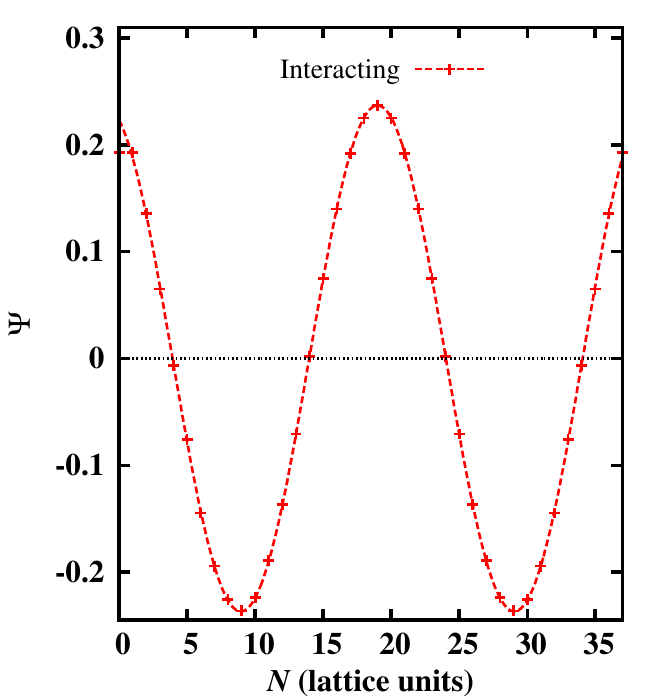}
\hfill
\includegraphics[angle=0,width=0.57\textwidth]{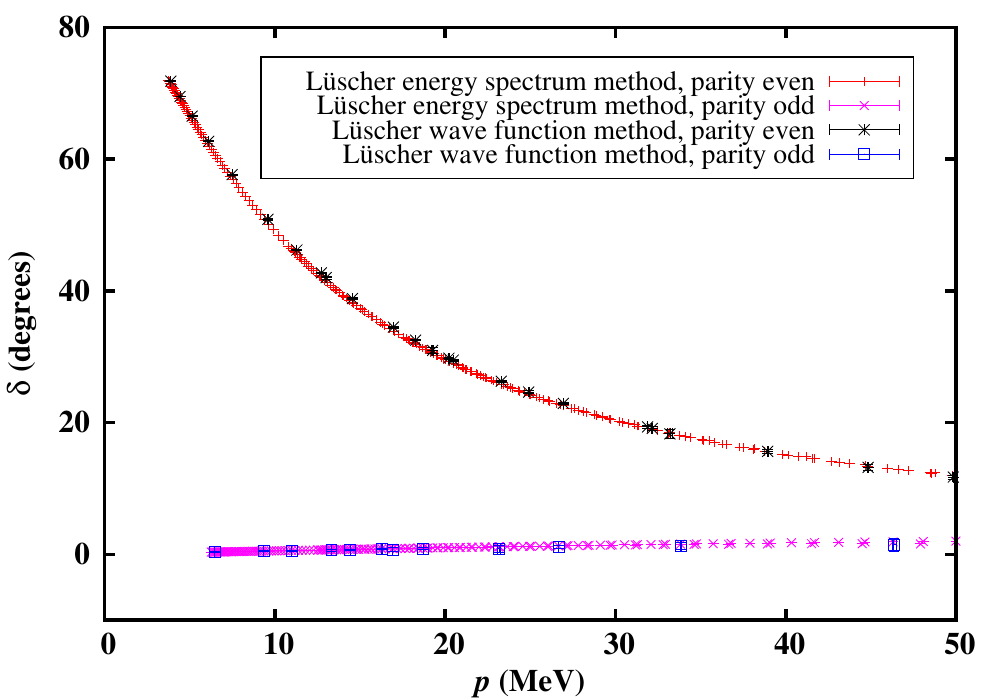}    
\caption{(Color online) Particle-dimer phase shifts  in one dimension calculated using the
  L\"{u}scher wave function method.
Left panel: An example of the wave function matching. Right panel:
Comparison of the phase shifts calculated using the L\"{u}scher periodic-box wave
function method and L\"{u}scher's finite-volume method with the exact energy spectrum.}
\label{pic1}
\end{figure}

We make use of the simplicity of the one dimensional system to benchmark several different
techniques for obtaining phase shifts. In each case we construct the adiabatic Hamiltonian using a projection time of $\tau = 0.30$~MeV. L\"{u}scher's energy spectrum
method makes use of the energy of a scattering state and not its wave function. However we can also calculate phase
shifts by fitting the wave function in the periodic box to its asymptotic form
Eq.~(\ref{eqn:1D}). Here and in what follows, we refer to this method as
the L\"{u}scher wave function method. In Fig.~\ref{pic1}(a) we show an
example of such a fit. 
The resulting value of the phase shift in
this example is \mbox{$\delta_0(p)=-160.5\pm 0.3^\circ$} with
momentum $p=31.28\pm0.03$  MeV. We can compare the phase shifts
obtained with this approach to L\"{u}scher's method using the exact energy spectrum. Fig.~\ref{pic1}(b) shows L\"{u}scher's energy spectrum method
phase shifts obtained for a number of lattices with $L=6\ldots 100$ in
lattice units.  With the L\"{u}scher periodic-box wave function method, phase shifts are
calculated for $L=10, 20, \ldots 100$ in lattice units. As expected,
one observes very good agreement between the L\"{u}scher
energy spectrum method and the  L\"{u}scher wave function fit. While this method clearly
works very well for the one-dimensional system, we find that fitting
periodic-box wave functions is much more problematic and less accurate in three dimensions,
especially for $\ell \ge 1$.  This is most likely due to systematic errors arising from lattice spacing artifacts.

\begin{figure}[tb]
\includegraphics[angle=0,width=0.38\textwidth]{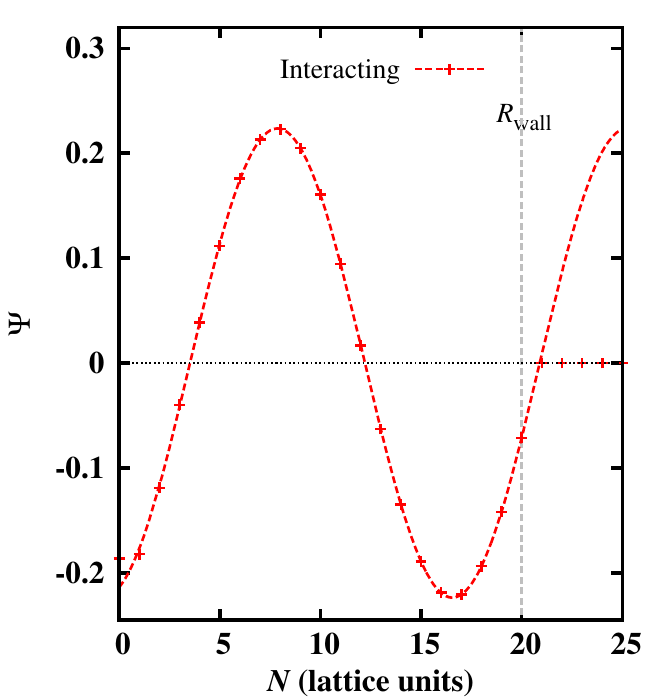}
\hfill
\includegraphics[angle=0,width=0.57\textwidth]{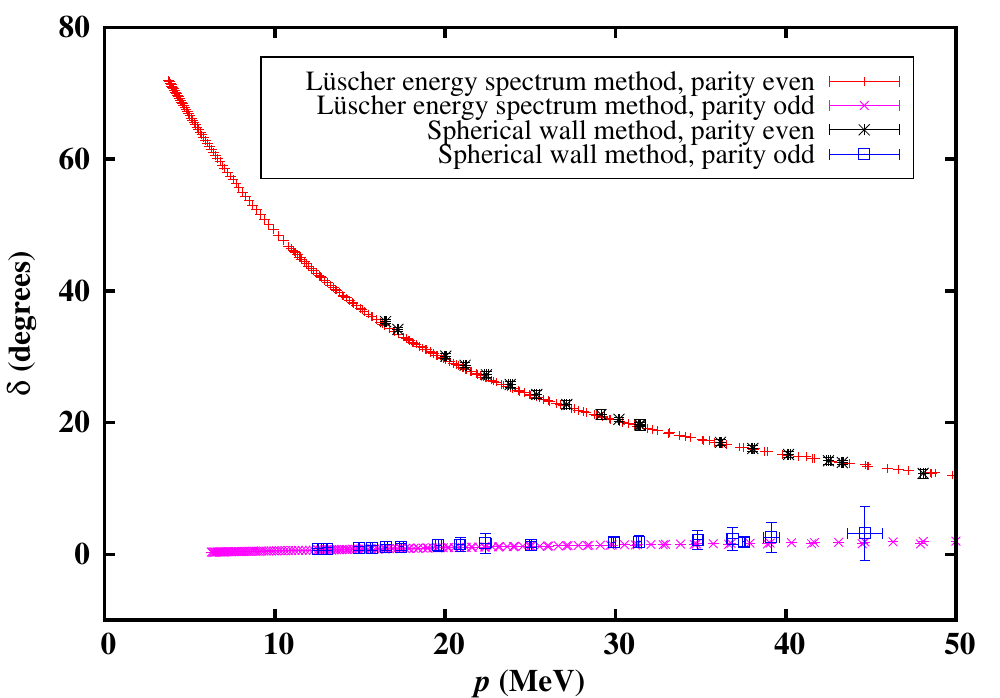}
\caption{(Color online) Particle-dimer phase shifts  in one dimension calculated using the spherical-wall method in
  one dimension with a three-parameter fit. 
Left panel: An example of the wave function fits. Right panel:
Comparison of the phase shifts calculated using the  spherical wall 
method and L\"{u}scher's finite-volume method with the exact energy spectrum. 
}
\label{pic2}
\end{figure}

The second method we consider is the spherical wall
approach described in the previous section. We impose a hard boundary on the relative separation and
calculate the phase shifts from the properties of the standing
wave functions. This method allows for the calculation of several data points
per chosen lattice volume, since we can vary the value of the wall radius, $R_{\rm{wall}}$. 
This represents an important computational advantage as compared to the L{\"u}scher wave
function method, especially for calculations in three dimensions. We will consider two versions of the spherical wall method.  In the first version we do a three parameter fit
of the overall normalization,
momentum and phase shift of the interacting wave functions. Fig.~\ref{pic2} shows the application of this approach for the case
$L=50$ and $R_{\rm{wall}}=20$ in lattice units. 
The resulting value of
the phase shift in this example is \mbox{$\delta_0(p)=-163.0\pm 0.4$}
for the momentum \mbox{$p=36.17\pm0.07$ MeV}. This method also shows very good agreement with L\"{u}scher's energy spectrum method. The phase shifts are calculated for $L=50$ and
$R_{\rm{wall}}=13\ldots 23$ in lattice units.

\begin{figure}[tb]
\includegraphics[angle=0,width=0.38\textwidth]{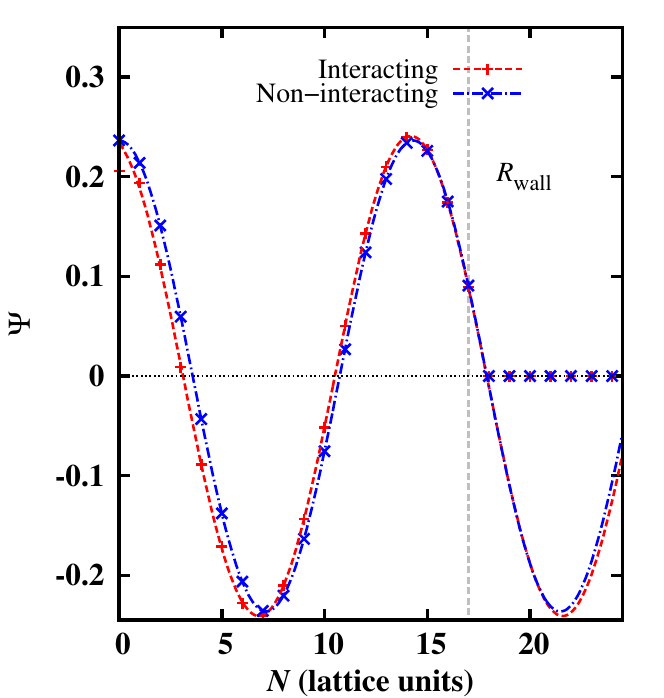}
\hfill
\includegraphics[angle=0,width=0.57\textwidth]{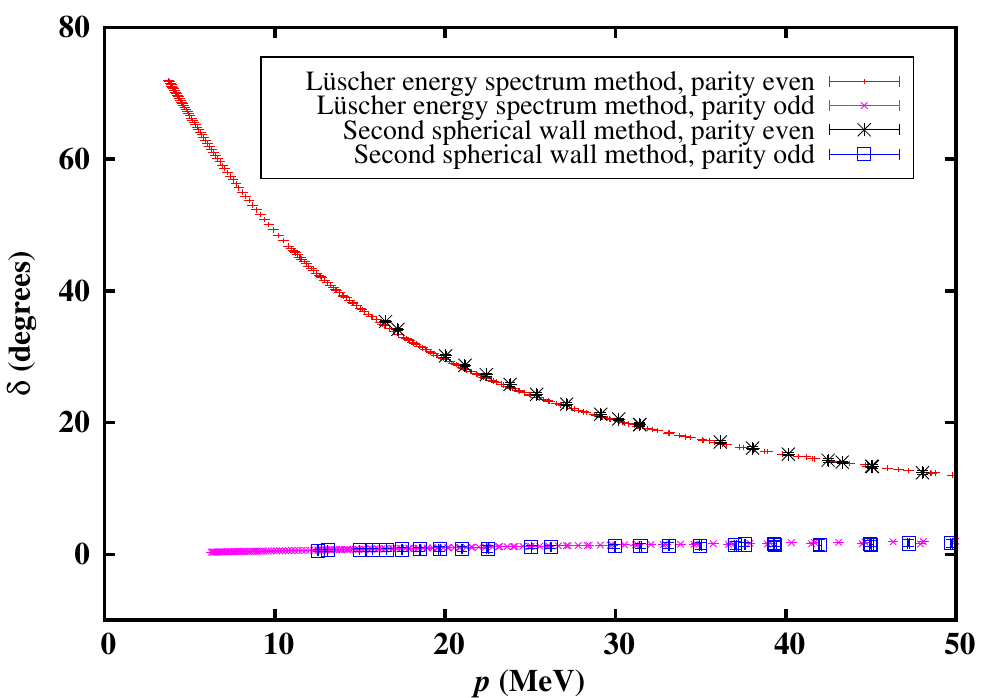}
\caption{(Color online) Particle-dimer phase shifts in one dimension using the second spherical
  wall approach with $R'_{\rm{wall}}$ determined from the non-interacting wave function.
Left panel: An example of the wave function fits. Right panel:
Comparison of the phase shifts calculated using the  second spherical
wall  approach 
and L\"{u}scher's finite-volume method with the exact energy spectrum. }
\label{pic3}
\end{figure}

Our next approach is a second version of the spherical wall wave function
method.  In this case we determine $R'_{\rm{wall}}=R_{\rm{wall}}+\epsilon$ from the non-interacting particle-dimer wave function. In the example shown
in Fig.~\ref{pic3}, the boundary is set at $R_{\rm{wall}}=17$, and we find the wave function vanishes at \mbox{$R'_{\rm{wall}}=17.901$}, and the momentum of the
free wave function is $p_0=43.874\pm0.02$~MeV. 
We then do a two-parameter fit to the interacting wave function and find
\mbox{$\delta_0(p)=-165.8\pm 0.5$} for the momentum
$p=42.5\pm0.1$~MeV.  The phase shifts shown in the right panel of Fig.~\ref{pic3} are
calculated for $L=50$ and $R_{\rm{wall}}=13\ldots 23$ in lattice
units. The results are in agreement with the results of the first spherical wall approach, but have smaller error bars, especially for the odd
parity phase shifts. In three dimensions this improvement becomes more
significant. 
 
\begin{figure}[tb]
\includegraphics[angle=0,width=0.38\textwidth]{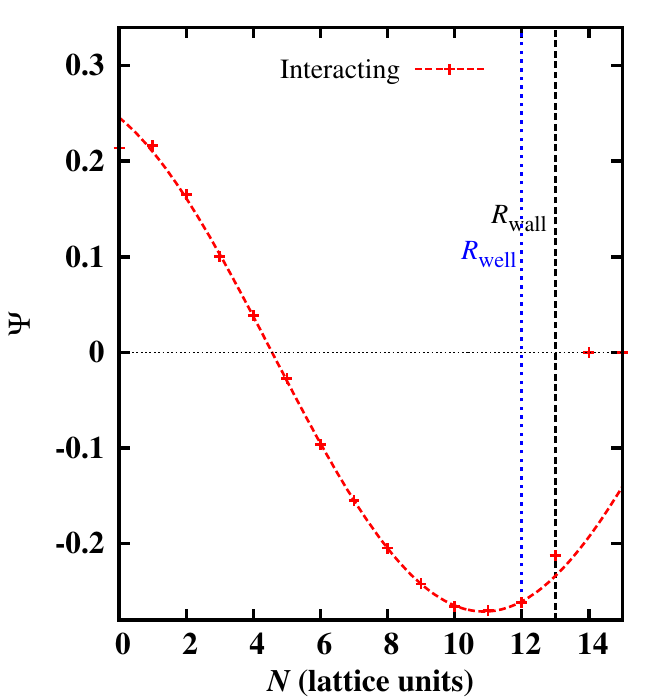}
\hfill
\includegraphics[angle=0,width=0.57\textwidth]{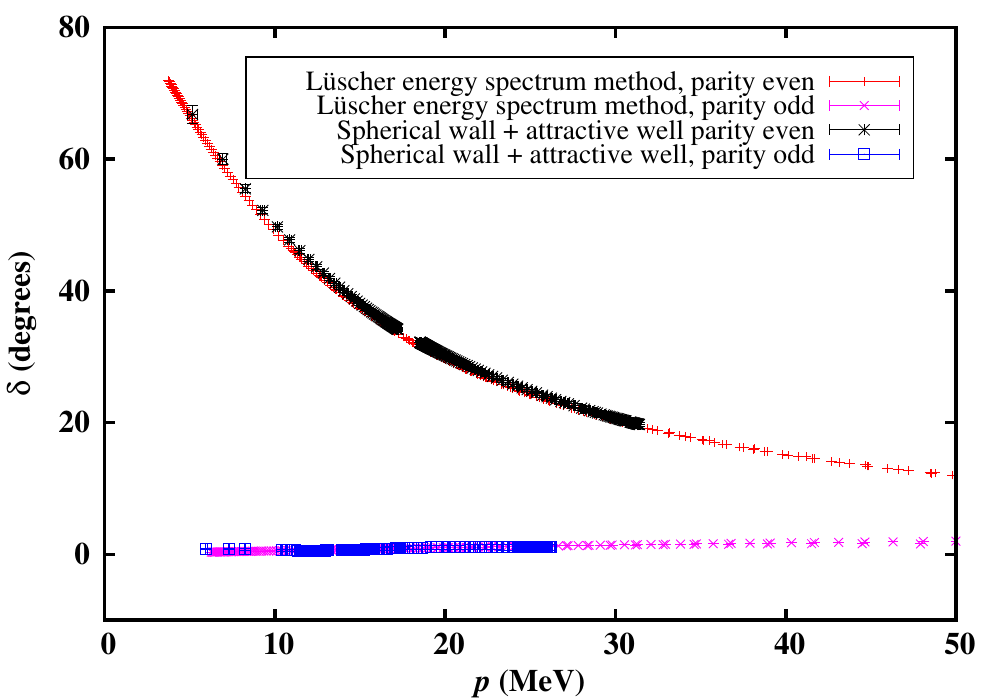}
\caption{(Color online) Particle-dimer phase shifts in one dimension with the spherical wall and an attractive well
  potential.
Left panel: An example of the wave function matching. Right panel:
Comparison of the  phase shifts calculated using the approach based on
the combination of the spherical wall and attractive well with
L\"{u}scher's finite-volume method using the exact energy spectrum. 
}
\label{pic4}
\end{figure}

One of the disadvantages of the spherical wall method is that one needs to go to rather large values of $R_{\rm{wall}}$ and $L$ in order to probe very low energies. The last method we consider overcomes this issue. In order
to compute phase shifts at low momenta using small lattices, we impose
a spherical hard wall and add also an attractive well potential in front of the wall boundary. We treat the depth of the well
as an adjustable continuous parameter. The
example shown in Fig.~\ref{pic4} corresponds to the case of $L=30$,
$R_{\rm{wall}}=13$ and $R_{\rm{well}}=12$ in 
lattice units. The resulting value of the phase shift in this example
is $\delta_0(p)=-155.0\pm 1.6^\circ$ at momentum $p=24.9\pm0.6$
MeV. The phase shifts shown in the right panel of  Fig.~\ref{pic4} are calculated for $L=50$,
$R_{\rm{wall}}=23$ and different well depths. 
The agreement with
L\"{u}scher's energy spectrum method is very good, and we also obtain
phase shifts for smaller momenta. However, the additional
attractive potential distorts the asymptotic form of the wave
function near the potential well. The distortion of the wave function grows with the depth of the
attractive potential, and the calculation of the phase shifts for
smaller momenta is achieved at the expense of a larger relative
error. This might complicate the application of this method   for calculations in three
dimensions where the values of $R_{\rm{wall}}$ are typically smaller.



\subsection{Fermion-dimer scattering in three dimensions}

Before presenting lattice calculations for fermion-dimer scattering in three dimensions, we first review continuum calculations of the same system in the limit of zero range interactions.  This corresponds to neutron-deuteron scattering at leading order
in  pionless effective field
theory~\cite{Bedaque:1999vb,Gabbiani:1999yv,Rupak:2001ci}. The $T$-matrix is obtained by solving  the Skorniakov-Ter-Martirosian (STM) integral equation,
\begin{align}
T_{\ell}(k,p) = - & \frac{8 \pi \gamma}{m p k}
\,
Q_{\ell}\left(\frac{p^{2}+k^{2}-mE-i0^{+}}{pk}\right)
\nonumber\\
&
-\frac{2}{\pi}\int_{0}^{\infty} dq \, \frac{q}{p} \frac{T_{\ell}(k,q)}{\sqrt{3q^{2}/4-mE-i0^{+}}-\gamma}
\,
Q_{\ell}\left(\frac{p^{2}+q^{2}-mE-i0^{+}}{pq}\right),
\label{eqn:t-matrix-STM}
\end{align}
where $\gamma$ is the dimer binding momentum, $E = 3p^{2}/(4m)-\gamma^{2}/m$ is the total energy, and $Q_{\ell}$ is the Legendre function of the second kind,
\begin{align}
Q_{\ell}(a) = \frac{1}{2}\int_{-1}^{1}
dx \, \frac{P_{\ell}(x)}{x+a}\,.
\end{align}
The scattering phase shift can then be extracted by matching the
solution of Eq.~(\ref{eqn:t-matrix-STM}) with the on-shell $T$-matrix 
\begin{align}
T_{\ell}(p,p) = \frac{3\pi}{m} \frac{p^{2\ell}}{p^{2\ell+1} \cot\delta_{\ell} - ip^{2\ell+1}}\,.
\end{align}
\begin{figure}[tb]
        \begin{center}
                {\includegraphics[width=3in]{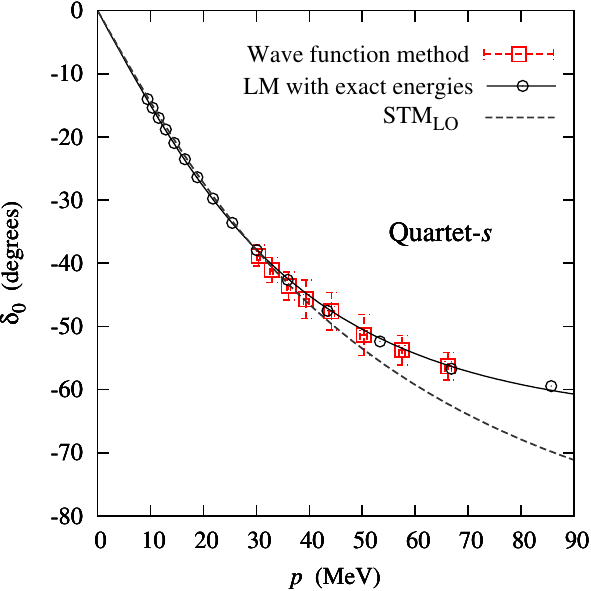}}
\hskip 0.5 true cm 
               {\includegraphics[width=3in]{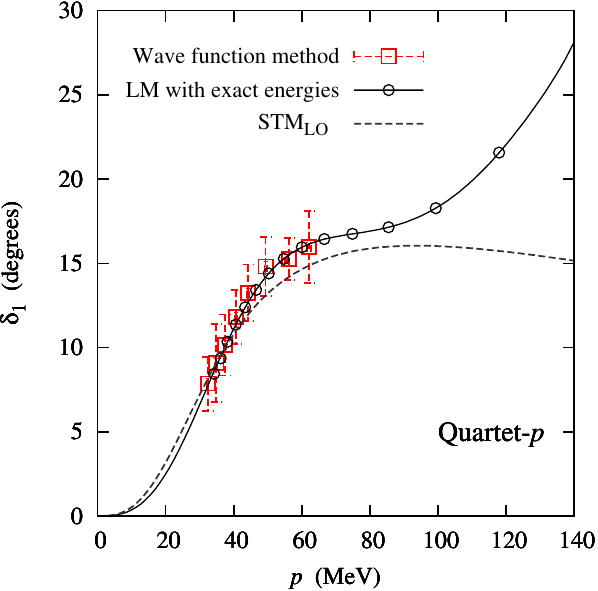}}
        \end{center}
        \caption{(Color online) The $s$-wave (left panel) and $p$-wave
          (right panel) scattering phase shifts for fermion-dimer scattering in three dimensions. We compare  phase shifts calculated using the the spherical
  wall approach with $R'_{\rm{wall}}$ determined from the non-interacting
wave function with
L\"{u}scher's finite-volume method using the exact energy spectrum.}
        \label{fig:sp-wavePhaseshift}
\end{figure}
For the three-dimensional fermion-dimer calculation, we use only the most promising of the four approaches explored previously for the one dimensional system. This is the spherical
wall method with $R'_{\rm{wall}}$ determined from the non-interacting wave function. In Fig.~\ref{fig:sp-wavePhaseshift}, the squares show lattice results for the $s$-wave and $p$-wave scattering phase
shifts using adiabatic projection method with $\tau =0.37$ and spherical wall method.  The circles are the exact lattice
results obtained using L\"uscher's method applied to  the  energies of the microscopic Hamiltonian $H$ evaluated using Lanczos eigenvector iteration. The dashed lines
correspond to leading order pionless EFT continuum results obtained using the STM\ equation. The solid lines are fits of 
the lattice data using an effective range expansion,  
\begin{align}
  p^{2\ell+1} \cot\delta_{\ell}(p)
  = -\frac{1}{a_{\ell}}+\frac{1}{2}r_{\ell} \,  p^{2}+\mathcal{O}(p^{4}) \,.
 \label{eqn:append:phaseshift-001}
\end{align}
We note again that these exact Lanczos benchmark calculations using the energies
of $H$ are only possible in
small systems.  The adiabatic projection method is needed to probe much
larger
systems via lattice Monte Carlo.  We have seen in Table~\ref{table:spectrum_H_Ha}
that when we use L\"uscher's method to extract phase shifts from the adiabatic
Hamiltonian energies, the errors are as large as 50\% at  low energies.  In comparison with this, we observe much smaller error bars and excellent agreement between the adiabatic Hamiltonian results and the exact lattice phase shifts in Fig.~\ref{fig:sp-wavePhaseshift}.
We should mention that the discrepancies
between the lattice and continuum results at large momenta 
are nothing more than lattice spacing artifacts and would go away in the limit of vanishing lattice spacing.

\section{Summary and Conclusions}
\label{sec:Summ}

In this paper we have presented a new method for computing scattering
parameters directly from cluster wave functions. It allows us to bypass
computation of the energy spectrum and thus to avoid potentially large
errors in  calculating low-energy nuclear scattering and reactions using the
adiabatic projection method.  We showed that the adiabatic Hamiltonian in the asymptotic
region reduces to a simple cluster Hamiltonian in a position space basis. In this way we can extract scattering phase shifts directly from the scattering cluster wave functions.

We considered particle-dimer scattering in one dimension and fermion-dimer scattering in three dimensions. In the one
dimensional particle-dimer example, we explored various versions of the adiabatic projection
method using the cluster wave functions rather than the finite volume
energies to extract phase shifts. First, we presented a simple matching of
the adiabatic lattice wave function in periodic box to the asymptotic form in Eq.~(\ref{eqn:1D}).  We were able to accurately fit the wave function in
the asymptotic region, and the calculated phase shifts were in
good agreement 
with exact lattice phase shifts computed using L\"uscher's energy method and
the spectrum of the microscopic Hamiltonian. Next, we imposed a
spherical hard wall boundary on the relative separation and matched the
resulting wave function to Eq.~(\ref{eqn:1D}). This approach was
also found to agree accurately with  the exact lattice phase shifts. We considered two variants of this spherical wall approach and found more accurate results when we use non-interacting
cluster wave functions to determine $R'_{\rm{wall}}$, the radius where the wave function vanishes.  We also tried adding an
attractive potential well in
addition to the spherical hard wall boundary in order to be able to
continuously vary the scattering energy. However, we found that this
potential well distorts the wave function in the asymptotic region and
results in larger errors of the phase shifts.

We then  considered the dimer-fermion
system in three spatial dimensions, which also  corresponds
to neutron-deuteron 
scattering in the spin-quartet channel. For this calculation we use adiabatic projection and the spherical
wall method with $R'_{\rm{wall}}$ determined from the non-interacting wave
function.  We find that this method gives good results for the phase shifts which agree well with the exact lattice phase shifts determined from Lanczos calculations of the spectrum of the microscopic Hamiltonian $H$.  

In this paper we have shown that very-high-precision energy calculations are not needed for determining phase shifts, and one can instead use spherical wall boundaries to measure phase shifts directly from scattering cluster wave functions.  The methods we have presented here are specially designed to be immediately useful for large-scale calculations of cluster-cluster scattering using lattice Monte Carlo.  Since the writing of the original draft of this work, the methods described here have been combined with lattice Monte Carlo
simulations to produce the first {\it ab initio} calculation of alpha-alpha scattering \cite{Elhatisari:2015iga}.  The adiabatic projection method is  used to reduce an eight-body system of nucleons to a system of two alpha particles.  There has also been significant improvements on the extraction of lattice phase shifts using the spherical wall method with auxiliary potentials such as attractive wells and complex-valued potentials for particles with spin and partial-wave mixing \cite{Lu:2015riz}.

We are confident that more applications are possible which combine the adiabatic projection method and Monte Carlo methods for scattering and reactions over a diverse range of few- and many-body systems.  We note, for example, recent developments using impurity lattice Monte Carlo \cite{Elhatisari:2014lka,Bour:2015a}, which opens the possibility of {\it ab initio} calculations of impurity quasiparticle scattering in quantum many-body systems. 

\section*{Acknowledgments}

The authors thank U.-G.~Mei{\ss}ner for valuable comments on the
manuscript. This work was supported in part by the U.S. Department of Energy
grant DE-FG02-03ER41260 (D.L. and M.P.), U.S. Department of Education
GAANN Fellowship (M.P.), the Seventh Framework Programme of EU,
 the ERC project 259218 NUCLEAREFT (E.E. and A.R.) and the European
Community-Research Infrastructure Integrating Activity ``Study of
Strongly Interacting Matter'' (acronym HadronPhysics3,
Grant Agreement n. 283286).


\begin{thebibliography}{99}

\bibitem{Navratil:2010jn} 
  P.~Navratil, R.~Roth and S.~Quaglioni,
  Phys.\ Rev.\ C {\bf 82}, 034609 (2010)
  [arXiv:1007.0525 [nucl-th]].

\bibitem{Navratil:2011zs} 
  P.~Navratil and S.~Quaglioni,
  Phys.\ Rev.\ Lett.\  {\bf 108}, 042503 (2012)
  [arXiv:1110.0460 [nucl-th]].

\bibitem{Romero-Redondo:2014fya} 
  C.~Romero-Redondo, S.~Quaglioni, P.~Navratil and G.~Hupin,
  Phys.\ Rev.\ Lett.\  {\bf 113}, 032503 (2014)
  [arXiv:1404.1960 [nucl-th]].


\bibitem{Neff:2010nm} 
  T.~Neff,
  Phys.\ Rev.\ Lett.\  {\bf 106}, 042502 (2011)
  [arXiv:1011.2869 [nucl-th]].

\bibitem{Neff:2010uk} 
  T.~Neff, H.~Feldmeier and K.~Langanke,
  Prog.\ Part.\ Nucl.\ Phys.\  {\bf 66}, 341 (2011)
  [arXiv:1011.2341 [nucl-th]].

\bibitem{Hagen:2013nca} 
  G.~Hagen, T.~Papenbrock, M.~Hjorth-Jensen and D.~J.~Dean,
  Rept.\ Prog.\ Phys.\  {\bf 77}, no. 9, 096302 (2014)
  [arXiv:1312.7872 [nucl-th]].

\bibitem{Nollett:2011qf} 
  K.~M.~Nollett and R.~B.~Wiringa,
  Phys.\ Rev.\ C {\bf 83}, 041001 (2011)
  [arXiv:1102.1787 [nucl-th]].

\bibitem{Carlson:2014vla} 
  J.~Carlson, S.~Gandolfi, F.~Pederiva, S.~C.~Pieper, R.~Schiavilla, K.~E.~Schmidt and R.~B.~Wiringa,
  arXiv:1412.3081 [nucl-th].

\bibitem{Liu:2005kr} 
  C.~Liu, X.~Feng and S.~He,
  Int.\ J.\ Mod.\ Phys.\ A {\bf 21}, 847 (2006)
  [hep-lat/0508022].

\bibitem{Lage:2009zv} 
  M.~Lage, U.-G.~Mei{\ss}ner and A.~Rusetsky,
  Phys.\ Lett.\ B {\bf 681}, 439 (2009)
  [arXiv:0905.0069 [hep-lat]].


\bibitem{Briceno:2013lba} 
  R.~A.~Briceno, Z.~Davoudi and T.~C.~Luu,
  Phys.\ Rev.\ D {\bf 88}, no. 3, 034502 (2013)
  [arXiv:1305.4903 [hep-lat]].

\bibitem{Briceno:2012yi} 
  R.~A.~Briceno and Z.~Davoudi,
  Phys.\ Rev.\ D {\bf 88}, no. 9, 094507 (2013)
  [arXiv:1204.1110 [hep-lat]].

\bibitem{Briceno:2013bda} 
  R.~A.~Briceno, Z.~Davoudi, T.~C.~Luu and M.~J.~Savage,
  Phys.\ Rev.\ D {\bf 88}, no. 11, 114507 (2013)
  [arXiv:1309.3556 [hep-lat]].

\bibitem{Briceno:2014oea} 
  R.~A.~Briceno,
  Phys.\ Rev.\ D {\bf 89}, no. 7, 074507 (2014)
  [arXiv:1401.3312 [hep-lat]].

\bibitem{Doring:2012eu} 
  M.~D\"oring, U.-G.~Mei{\ss}ner, E.~Oset and A.~Rusetsky,
  Eur.\ Phys.\ J.\ A {\bf 48}, 114 (2012)
  [arXiv:1205.4838 [hep-lat]].


\bibitem{Polejaeva:2012ut} 
  K.~Polejaeva and A.~Rusetsky,
  Eur.\ Phys.\ J.\ A {\bf 48}, 67 (2012)
  [arXiv:1203.1241 [hep-lat]].

\bibitem{Briceno:2012rv} 
  R.~A.~Briceno and Z.~Davoudi,
  Phys.\ Rev.\ D {\bf 87}, no. 9, 094507 (2013)
  [arXiv:1212.3398 [hep-lat]].

\bibitem{Meissner:2014dea} 
  U.-G.~Mei{\ss}ner, G.~Ri­os and A.~Rusetsky,
  Phys.\ Rev.\ Lett.\  {\bf 114}, no. 9, 091602 (2015)
  [arXiv:1412.4969 [hep-lat]].

\bibitem{Epelbaum:2011md} 
  E.~Epelbaum, H.~Krebs, D.~Lee and U.-G.~Mei{\ss}ner,
  Phys.\ Rev.\ Lett.\  {\bf 106}, 192501 (2011)
  [arXiv:1101.2547 [nucl-th]].

\bibitem{Epelbaum:2012qn} 
  E.~Epelbaum, H.~Krebs, T.~A.~L\"ahde, D.~Lee and U.-G.~Mei{\ss}ner,
  Phys.\ Rev.\ Lett.\  {\bf 109}, 252501 (2012)
  [arXiv:1208.1328 [nucl-th]].


\bibitem{Lahde:2013uqa} 
  T.~A.~L\"ahde, E.~Epelbaum, H.~Krebs, D.~Lee, U.-G.~Mei{\ss}ner and G.~Rupak,
  Phys.\ Lett.\ B {\bf 732}, 110 (2014)
  [arXiv:1311.0477 [nucl-th]].


\bibitem{Epelbaum:2013paa} 
  E.~Epelbaum, H.~Krebs, T.~A.~L\"ahde, D.~Lee, U.-G.~Mei{\ss}ner and G.~Rupak,
  Phys.\ Rev.\ Lett.\  {\bf 112}, no. 10, 102501 (2014)
  [arXiv:1312.7703 [nucl-th]].

\bibitem{Rupak:2013aue} 
  G.~Rupak and D.~Lee,
  Phys.\ Rev.\ Lett.\  {\bf 111}, no. 3, 032502 (2013)
  [arXiv:1302.4158 [nucl-th]].


\bibitem{Pine:2013zja} 
  M.~Pine, D.~Lee and G.~Rupak,
  Eur.\ Phys.\ J.\ A {\bf 49}, 151 (2013)
  [arXiv:1309.2616 [nucl-th]].

\bibitem{Elhatisari:2014lka} 
  S.~Elhatisari and D.~Lee,
  Phys.\ Rev.\ C {\bf 90}, no. 6, 064001 (2014)
  [arXiv:1407.2784 [nucl-th]].


\bibitem{Luscher:1986pf} 
  M.~L\"uscher,
  Commun.\ Math.\ Phys.\  {\bf 105}, 153 (1986).

\bibitem{Luscher:1990ux} 
  M.~L\"uscher,
  Nucl.\ Phys.\ B {\bf 354}, 531 (1991).

\bibitem{Luu:2011ep} 
  T.~Luu and M.~J.~Savage,
  Phys.\ Rev.\ D {\bf 83}, 114508 (2011)
  [arXiv:1101.3347 [hep-lat]].

\bibitem{Fu:2011xz} 
  Z.~Fu,
  Phys.\ Rev.\ D {\bf 85}, 014506 (2012)
  [arXiv:1110.0319 [hep-lat]].

\bibitem{Leskovec:2012gb} 
  L.~Leskovec and S.~Prelovsek,
  Phys.\ Rev.\ D {\bf 85}, 114507 (2012)
  [arXiv:1202.2145 [hep-lat]].

\bibitem{Gockeler:2012yj} 
  M.~G\"ockeler, R.~Horsley, M.~Lage, U.-G.~Mei{\ss}ner, P.~E.~L.~Rakow, A.~Rusetsky, G.~Schierholz and J.~M.~Zanotti,
  Phys.\ Rev.\ D {\bf 86}, 094513 (2012)
  [arXiv:1206.4141 [hep-lat]].

\bibitem{Luscher:1985dn} 
  M.~L\"uscher,
  Commun.\ Math.\ Phys.\  {\bf 104}, 177 (1986).

\bibitem{Konig:2011nz} 
  S.~K\"onig, D.~Lee and H.-W.~Hammer,
  Phys.\ Rev.\ Lett.\  {\bf 107}, 112001 (2011)
  [arXiv:1103.4468 [hep-lat]].

\bibitem{Konig:2011ti} 
  S.~K\"onig, D.~Lee and H.-W.~Hammer,
  Annals Phys.\  {\bf 327}, 1450 (2012)
  [arXiv:1109.4577 [hep-lat]].

\bibitem{Bour:2011ef} 
  S.~Bour, S.~K\"onig, D.~Lee, H.-W.~Hammer and U.-G.~Mei{\ss}ner,
  Phys.\ Rev.\ D {\bf 84}, 091503 (2011)
  [arXiv:1107.1272 [nucl-th]].

\bibitem{Bour:2012hn} 
  S.~Bour, H.-W.~Hammer, D.~Lee and U.-G.~Mei{\ss}ner,
  Phys.\ Rev.\ C {\bf 86}, 034003 (2012)
  [arXiv:1206.1765 [nucl-th]].

\bibitem{Borasoy:2007vy} 
  B.~Borasoy, E.~Epelbaum, H.~Krebs, D.~Lee and U.-G.~Mei{\ss}ner,
  Eur.\ Phys.\ J.\ A {\bf 34}, 185 (2007)
  [arXiv:0708.1780 [nucl-th]].

\bibitem{Carlson:1984zz} 
  J.~Carlson, V.~R.~Pandharipande and R.~B.~Wiringa,
  Nucl.\ Phys.\ A {\bf 424}, 47 (1984).

\bibitem{Lu:2015gfa} 
  B.~N.~Lu, T.~A.~L{\"a}hde, D.~Lee and U.-G.~Mei{\ss}ner,
  arXiv:1504.01685 [nucl-th].
 
\bibitem{Lu:2014xfa} 
  B.~N.~Lu, T.~A.~L{\"a}hde, D.~Lee and U.-G.~Mei{\ss}ner,
  Phys.\ Rev.\ D {\bf 90}, no. 3, 034507 (2014)
  [arXiv:1403.8056 [nucl-th]].
  
\bibitem{Johnson:1982yq} 
  R.~C.~Johnson,
  Phys.\ Lett.\ B {\bf 114}, 147 (1982).

\bibitem{Bedaque:1999vb} 
  P.~F.~Bedaque and H.~W.~Grie{\ss}hammer,
  Nucl.\ Phys.\ A {\bf 671}, 357 (2000)
  [nucl-th/9907077].

\bibitem{Gabbiani:1999yv} 
  F.~Gabbiani, P.~F.~Bedaque and H.~W.~Grie{\ss}hammer,
  Nucl.\ Phys.\ A {\bf 675}, 601 (2000)
  [nucl-th/9911034].

\bibitem{Rupak:2001ci} 
  G.~Rupak and X.~W.~Kong,
  Nucl.\ Phys.\ A {\bf 717}, 73 (2003)
  [nucl-th/0108059].

\bibitem{Elhatisari:2015iga} 
  S.~Elhatisari, D.~Lee, G.~Rupak, E.~Epelbaum, H.~Krebs, T.~A.~L{\"a}hde, T.~Luu and U.-G.~Mei{\ss}ner,
  arXiv:1506.03513 [nucl-th].
  
\bibitem{Lu:2015riz} 
  B.~N.~Lu, T.~A.~L{\"a}hde, D.~Lee and U.-G.~Mei{\ss}ner,
  arXiv:1506.05652 [nucl-th].

\bibitem{Bour:2015a}
S.~Bour, D.~Lee, H.-W.~Hammer and U.-G.~Mei{\ss}ner,
arXiv:1412.8175 [cond-mat.quant-gas].


  
\end{thebibliography}

\end{document}